\documentclass[aps,prl,twocolumn,amsfonts,showpacs]{revtex4} 
\usepackage{bm}
\usepackage{epsfig,amsopn}
\usepackage{graphicx}
\usepackage{amsmath,amssymb}
\usepackage{natbib}
\newcommand\bea{\begin{eqnarray}}
\newcommand\eea{\end{eqnarray}}
\newcommand\beq{\begin{equation}}
\newcommand\eeq{\end{equation}}

\newcommand{\bib}{\bibitem}

\def\nn{\nonumber}
\def\dg{\dagger}
\def\f{\frac}
\def\la{\langle}
\def\ra{\rangle}

\def\b{\beta}

\def\e{\epsilon}
\def\g{\gamma}

\def\s{\sigma}

\def\ua{\uparrow}
\def\da{\downarrow}
\def\bk{{\bf k}}
\def\bkn{{\bf k}_{N}}
\def\bq{{\bf q}}
\def\bqn{{\bf q}_{N}}
\def\bl{{\bf x}}
\def\b0{{\bf 0}}

\begin{document}

\title{Scattering of electrons from an interacting region}
\author{Abhishek Dhar$^1$, Diptiman Sen$^2$ and Dibyendu Roy$^1$}
\affiliation{$^1$ Raman Research Institute, Bangalore 560080, India \\
$^2$ Centre for High Energy Physics, Indian Institute of Science,
Bangalore 560012, India}

\begin{abstract}
We address the problem of transmission of electrons between two 
noninteracting leads through a region where they interact (quantum dot). We
use a model of spinless electrons hopping on a one-dimensional lattice and 
with an interaction on a single bond. We show that all the two-particle 
scattering states can be found exactly. Comparisons are made with 
numerical results on the time evolution of a two-particle wave packet and 
several interesting features are found. 
For $N$ particles the scattering state is obtained within a two-particle
scattering approximation. 
For a dot connected to Fermi seas at different chemical potentials, we find 
an expression for the change in the Landauer current resulting from the 
interactions on the dot. We end with some comments on the case of spin-1/2 
electrons.
\end{abstract}
\vskip .5 true cm

\pacs{73.21.Hb, 03.65.Nk, 73.50.Bk}
\maketitle


An understanding of the behavior of electrons interacting with each other in 
a localized region has been a challenging problem in theoretical physics. 
Recently it has attracted much attention in view of the experimental interest 
in transport across quantum dots and the Kondo effect in a quantum dot 
\cite{kondo1}. As a prototypical model, let us consider two ideal leads, 
where all electronic interactions can be neglected, connected to a region (a 
quantum dot) where the electrons interact. One is interested in the current 
through the dot in response to an applied voltage difference between the leads.

As has been discussed in Ref. \cite{mehta06}, there are several different
but equivalent theoretical approaches. In the 
nonequilibrium Green's function (NEGF) approach the initial density matrix, 
of the two reservoirs (taken as ideal Fermi liquids in equilibrium at different
chemical potentials) and the dot (in an arbitrary initial state), is evolved 
in time. The coupling between the reservoirs and the dot is switched on
adiabatically and one looks at steady state properties of the resulting 
density matrix. A related approach is the quantum Langevin method where the 
reservoirs are treated as sources of noise and dissipation. 
A second approach is to view this as a time-independent 
scattering problem and to look for many-particle scattering states which
have the correct asymptotic form in the leads. This is in the spirit of the 
Landauer formalism. In the case where there are 
no interactions in the dot region, exact results for the current and other 
steady state properties can be obtained, and all three approaches give 
identical answers \cite{caroli71,todorov93,wingreen,dharsen06}.

The interacting case however is much more difficult to study. For a single 
dot connected to noninteracting leads, some results using the NEGF method have
been obtained using the so-called non-crossing approximation 
\cite{wingreen}. For an integrable model, namely the interacting resonance 
level model, Mehta and Andrei used the scattering approach to solve the 
problem exactly \cite{mehta06}. Using the Bethe ansatz, they were able to 
express all $N$-particle scattering states in terms of the two-particle 
$S$-matrix which is known exactly. They considered a continuum model with a 
linear spectrum which makes it integrable. The $N$-particle scattering matrix 
for electrons interacting in a quantum dot has also been studied in 
Ref. \cite{lebedev,goorden08}.

In this Letter, we study a lattice version of the model considered in Ref.
\cite{mehta06}. We show here that using the Lippman-Schwinger method all 
two-particle eigenstates of this model can be found {\it exactly.} The form 
of the $S$-matrix indicates that the model is not solvable by the Bethe 
ansatz. We examine the $S$-matrix and compare it with numerical experiments 
on scattering of a two-particle wave packet. We also study many-body transport
in this system by considering $N$-particle states corresponding to left and 
right leads with different chemical potentials. We obtain an expression for 
the change in the Landauer current arising from the interactions.

We note that the study of two-particle scattering states is in itself of 
interest \cite{aharony1,aharony2}, apart from being the starting point for 
the study of many-particle
states necessary to understand transport. Recently, Goorden and B\"uttiker
\cite{goorden07} have studied a set-up with two disconnected
conducting wires and with electrons in the two wires interacting weakly in a 
localized region. Using first order perturbation theory, the two-particle 
$S$-matrix was evaluated and used to extract information on transmission and 
correlations in a two-particle scattering experiment. In our single channel 
case, we will show that the antisymmetry of the wave functions leads to 
striking asymmetries in the $S$-matrix. In another interesting recent
work, the $S$-matrix in a model of two photons interacting with a 
localized atom was studied \cite{shen}. 

We consider a tight-binding one-dimensional lattice with spinless
electrons. The model considered describes an interacting dot on the
sites $x=0,1$ which is connected to two noninteracting one-dimensional 
leads on either side. The Hamiltonian is given by
\bea H ~&=&~H_L ~+~ H_D ~+~ V_C, ~~~{\rm where} \label{ham} \\
H_L &=& - \sum_{x=-\infty}^{\infty} \hspace{-0.15cm}' ~~(c_x^\dg c_{x+1} + 
c_{x+1}^\dg c_x ), \nn \\
H_D &=& -\g (c_0^\dg c_1+c_1^\dg c_0) + e (n_0 + n_1) + U n_0 n_1, \nn \\
{\rm and}~~ V_C&=&- \g' (c^\dg_{-1}c_0+c_0^\dg c_{-1}+c^\dg_1 c_2+c_2^\dg c_1),
\nn \eea
where $n_x=c_x^\dg c_x$ is the number operator at site $x$, and $\sum'$
implies omission of $x=-1,0,1$ from the summation. We set the lattice spacing 
and $\hbar$ to $1$. In this paper we only consider the case $\g = \g' = 1$ 
and $e = 0$ corresponding (for $U=0$) to the case of a perfectly transmitting 
dot but the general case can be treated similarly \cite{sen}. 

{\bf Scattering states:}~
We first show how one can obtain all the two-particle energy 
eigenstates exactly for this problem. Consider the 
noninteracting Hamiltonian $H_0=H$ with $U=0$. For this case, the 
one-particle eigenstates have the form $\phi_k (x) = e^{ikx}$
with energy $E_k = - 2 \cos k$, where $-\pi < k \le \pi$. Now consider a 
two-particle incoming state given by $\phi_{\bk} (\bl)= e^{i(k_1 x_1 + 
k_2 x_2)} - e^{i(k_2 x_1 + k_1 x_2)}$, with $\bk=(k_1,k_2)$
and $\bl=(x_1,x_2)$. The energy of this state is $E_{\bk}=E_{k_1}+E_{k_2}$. 
A scattering eigenstate $|\psi\ra$ of $H=H_0+V$ (where $V=U n_0 n_1$) with 
energy $E$ is related to a 
state $|\phi\ra$ of $H_0$ by the Lippman-Schwinger equation
\bea |\psi\ra &=& |\phi\ra ~+~ G_0^+(E) V |\psi\ra, \label{lipp} \eea 
where $G_0^+(E) = {1}/{(E- H_0 +i \e)}$.
In the two-particle sector, in the position basis $|\bl \ra$ and with an
incident state $<\bl|\phi \ra=\phi_\bk(\bl)$, Eq. (\ref{lipp}) gives
\bea \psi_\bk(\bl) &=& \phi_\bk(\bl) ~+~ UK_{E_\bk}(\bl)~ \psi_\bk (\b0), 
\label{scatt} \eea
where $K_{E_\bk}(\bl) = \la \bl| G_0^+(E_\bk) |\b0 \ra$ and $\b0 \equiv 
(1,0)$. We can determine $\psi_\bk(\b0)$ using Eq. (\ref{scatt}), 
$\psi_\bk(\b0)$$={\phi_\bk(\b0)}/[{1-UK_{E_\bk}(\b0)}]$.
The matrix element $K_{E_\bk}(\bl)$ is explicitly given by
\bea K_{E_\bk}(\bl) = g^+_{E_\bk}(x_1-1,x_2) ~-~ g^+_{E_\bk} (x_1,x_2-1), 
\label{G0eq} \eea
where $g^+_{E_\bk}(\bl) =[1/(2 \pi)^2] \int_{-\pi}^\pi \int_{-\pi}^\pi 
{dq_1 dq_2} {e^{i {\bf q} \cdot \bl }}/ (~E_\bk - E_{\bf q} +i \e)$ 
is the usual two-dimensional lattice Green's function. It is instructive to 
look at the asymptotic form of the scattered wave function \cite{economou};
this can be obtained by the saddle point method, the contribution to the 
integral in Eq. (\ref{G0eq}) coming from the region near $E_{\bf q} = E_\bk$.
Apart from a factor $U\psi_\bk(\b0)$, we find asymptotically that
\bea K^{as}_{E_\bk}(\bl) &=& \f{(\pm 1-i)}{4 \pi^{1/2}} \f{e^{i( k'_1 x_1+
 k'_2 x_2)}}{(r/r_0)^{1/2}} (e^{-i k'_1}-e^{-i k'_2}), \nn \\
& & \label{asymwf1} \\
{\rm with}~~ \f{x_1 }{\sin(k'_1)} &=& \f{x_2} {\sin(k'_2)}, ~~{\rm where} ~~
x_i /\sin(k'_i) > 0, \label{cond1} \\
E_\bk &=& -~ 2\cos(k'_1) ~-~ 2\cos(k'_2), \label{cond2} \\
r &=& (x_1^2+x_2^2)^{1/2}, \label{rll} \\
{\rm and}~~ r_0&=&\f{[\sin^2(k'_1)+\sin^2(k'_2)]^{1/2}}
{|\sin^2(k'_1)\cos(k'_2)+\sin^2(k'_2)\cos(k'_1)|}, \nn \eea
where the $\pm$ sign in Eq. (\ref{asymwf1}) corresponds to $E_\bk \gtrless 0$.
The antisymmetry of the wave function is implicitly hidden in the
$\bl$-dependence of $\bk'$. [The expression in Eq. (\ref{asymwf1})
is clearly more complicated than the Bethe ansatz would have given
which is a superposition of only four pairs of momenta, namely, $(\pm k_1,
\pm k_2)$.] The physical interpretation of the above solution is as follows.
Two electrons with initial momenta $(k_1,k_2)$ emerge, after scattering, with 
momenta $(k'_1,k'_2)$. Energy is conserved as implied by Eq. (\ref{cond2}). 
(Momentum is not conserved because the interaction term $Un_0 n_1$
breaks translation invariance). The velocities of the electrons are given by
$v_1=2 \sin(k'_1)$ and $v_2=2\sin(k'_2)$; Eq. (\ref{cond1}) expresses the 
fact that the electrons observed at $(x_1,x_2)$ must reach there at the same 
time after collision. Note that we can equivalently think of this problem as 
that of a single electron in a two-dimensional ($2D$) lattice moving in the
half-space $x_1>x_2$, with a hard wall along the diagonal $x_1=x_2$
and a single impurity at the site $\b0$. The particle 
flux ${\vec{J}} \cdot d{\vec{S}}$ in a given direction $tan(\theta)=x_2/x_1$ 
in the $2D$ problem corresponds, in the $1D$ problem, to the rate at which two
particles are scattered with velocity ratio $v_2/v_1=tan(\theta)$.
Instead of the usual scattering cross-section, it is useful here to
calculate the scattering rate for {\emph{unit two-particle density at
the site $\b0$}}. This is given by
\bea && |f(\theta)|^2 ~d \theta ~=~ \f{{\vec{J}} \cdot d{\vec{S}}}{|\phi_\bk
(\b0)|^2} ~=~ \f{1}{|1/U-K_{E_\bk}(\b0)|^2} \nn \\
&& ~~\times\f{[1-\cos(k'_1-k'_2)]~[\sin^2(k'_1)+\sin^2(k'_2)]} 
{2 \pi |\sin^2(k'_1)\cos(k'_2)+\sin^2(k'_2) \cos(k'_1)|} ~d \theta, \nn \\
&& \label{scatform1} \eea
where $k'_1,k'_2$ are known in terms of $\theta$. Experimentally it
may be simpler to find the number of particles scattering within an
energy interval $dE_{k'_2}$ (energy conservation implies that $dE_{k'_1} +
dE_{k'_2} = 0$). Defining $P(E_{k_1}, E_{k_2} \to E_{k'_1},E_{k'_2}) 
dE_{k'_2} = |\phi_\bk (\b0)|^2 |f(\theta)|^2 d \theta$, we find that
$P = {[1-\cos(k_1-k_2)]}{[1-\cos(k'_1-k'_2)]}/\{ |1/U-K_{E_\bk}(\b0)|^2~ 2 
\pi |\sin (k'_1) \sin (k'_2)| \}$.

\begin{figure}
\vspace{1cm}
\includegraphics[width=3in]{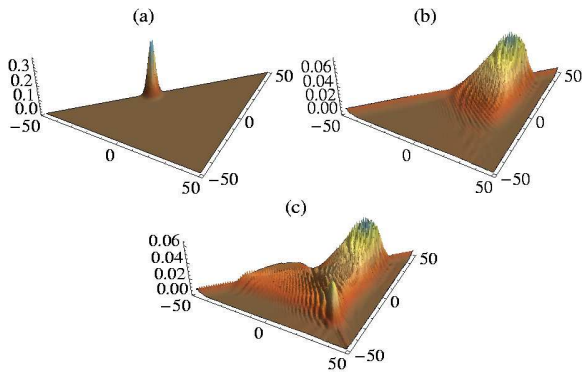}
\caption{Plot of the evolution of an incident wave packet (a) after passing 
through the origin with $U=0$ in (b) and $U=2$ in (c). Note the strong 
scattering at an angle $\theta=-\pi/4$.} \label{3dplot} \end{figure}

\begin{figure}
\vspace{1cm}
\includegraphics[width=3in]{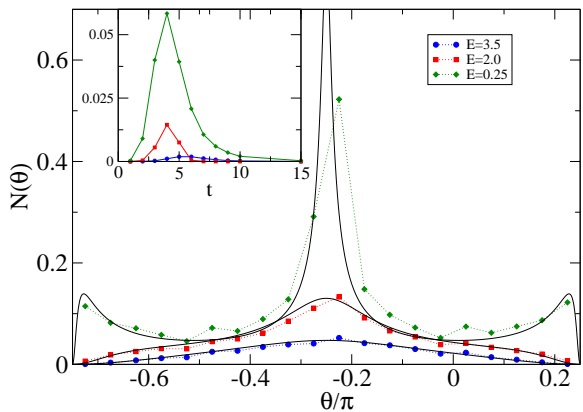}
\caption{Plot of the number of particles scattered into a given direction for 
incident wave packets with different energies and $U=2$. The bold lines show 
the results from scattering theory estimated using $|f(\theta)|^2$ and the 
incident particle density at the origin (inset). Inset shows 
$|\Psi_{inc}(\b0,t)|^2$.} \label{fvstheta} \end{figure}

For the two-particle case it is more useful to study wave packets. 
We now consider the time evolution of wave packets and
see how well the predictions of the scattering theory hold. The
scattering states given by Eq. (\ref{scatt}) are the full set of allowed 
two-particle energy eigenstates (for $U > \pi$ one gets an additional
bound state \cite{sen}). These can be generated by a unitary time
evolution of the unperturbed states which form a complete set. Hence
these states also form a complete set,
and any two-particle wave function can be expanded using this basis. Thus 
the time evolution of an initial wave packet $\Psi(\bl,t=0)$ is given by
\bea & & \Psi(\bl,t) = \f{1}{(2 \pi)^2} \int_{-\pi}^\pi dq_1\int_{-\pi}^{q_1}
dq_2~ a({\bf q}) \psi_\bq(\bl) ~e^{-iE_\bq t}, \nn \\ 
& & {\rm where} ~~a(\bq) = \sum_{x_1>x_2} \Psi(\bl,t=0) ~\psi^*_\bq (\bl). \eea
The time evolution can be studied quite accurately because of our
knowledge of the exact basis states. In evaluating the basis states,
for small $(x_1,x_2) \lesssim 15$, we evaluate the Green's functions 
$g^+_{E_\bk}(\bl)$ exactly using recursion relations 
\cite{morita71} relating these to $g_{E_\bk}^+(0,0)$ and
$g_{E_\bk}^+(1,1)$. For larger $(x_1,x_2)$ we use the asymptotic forms
which are quite accurate. We find that in our computations the
normalization of the wave function is preserved to within $0.5\%$. 
In Fig. \ref{3dplot} we show the typical time-evolution of a wave packet 
with initial position and momentum localized at $\bl=(-3.922,-5.0)$ and
$\bq=(2.554,0.785)$ respectively and with widths $\delta x \approx \delta q 
\approx 1$~ and $E \approx 0.25$. These initial conditions have been chosen
so that the two particles reach the site $\b0$ at roughly the same time;
this maximizes their interaction. The initial wave packet shown in Fig. 1 
($a$) evolves at time $t=20$ to ($b$) for $U=0$ and to ($c$) for $U=2$.
For the scattered wave function in Fig. \ref{3dplot} (c) we can count
the number of particles scattered into a given direction. This is
plotted in Fig. \ref{fvstheta} for incident wave packets with
different energies. We also compare this with the
scattering theory prediction by plotting $|f(\theta)|^2$ multiplied by
the time-integrated incident two-particle density at the origin. The
comparison can be seen to be very good.

{\bf Transport calculation:}
We will now turn our attention to quantities of interest in transport 
calculations. The current density is given by the expectation
value of the operator $j_x = -i(c_x^\dg c_{x+1}-h.c.)$ in the scattering state
$|\psi_\bk\ra=|\phi_\bk\ra +|S_\bk\ra$. The current in the incident state is 
given by $\la \phi_\bk| j_x |\phi_\bk \ra =2[\sin(k_1) +\sin (k_2)] {\cal N}$,
where ${\cal{N}}$ is the total number of sites in the entire system. The
change in current due to scattering, $\delta j(k_1, k_2) =\la \psi_\bk |j_x| 
\psi_\bk \ra - \la \phi_\bk | j_x | \phi_\bk \ra$, gets contributions from 
two parts, namely, $j_S
= \la S_\bk|j_x|S_\bk \ra$ and $j_C=\la S_\bk|j_x|\phi_\bk\ra +\la\phi_\bk
|j_x |S_\bk\ra$, and is of order $1$, i.e., it is a factor of $\cal N$ smaller
than the current in the incident state \cite{aharony2}. 
We find that
\beq \delta j (k_1, k_2) ~=~ \f{2 |\phi_\bk (\b0)|^2 Im [K_{E_\bk}(\b0)]}{
|1/U-K_{E_\bk} (\b0)|^2} ~[sgn (k_1) + sgn (k_2)], \label{c2} \eeq
where $sgn (k) \equiv |k|/k$.

{\emph {$N$-particle scattering state and change in the Landauer
current}:}~We now consider the problem of calculating the current in a 
situation where the interacting region is connected to left and right leads 
which are at zero temperature and chemical potentials $\mu_L$ and $\mu_R$
respectively. In that case we have to consider an initial state with
$N_L$ electrons in positive momentum states filling $1$-particle
energy levels up to $\mu_L$ and $N_R$ electrons in negative momentum
states filling levels up to $\mu_R$. Let $N=N_L+N_R$ and let us denote this 
$N$-particle incident wave by $|\phi^{(N)}\ra=|\bk^{(N)}\ra$, where 
$\bk^{(N)} =\{k_1 k_2...k_N\}$. One then needs to find the corresponding 
scattering state and compute the particle current. An exact solution 
for the $N$-particle scattering state looks difficult. We will therefore
restrict ourselves to an approximation in which only two-particle scattering
is taken into account. Within this approximation, the scattered wave is given 
by $|\psi_{\bkn} \ra= |\phi_{\bkn} \ra + |S_{\bkn}\ra$, where the transition 
amplitude to a wave vector $\bqn= \{q_1q_2...q_N\}$ is given by
\bea \la \bqn| S_{\bkn}\ra &=& \sum_{\bq_{2} \bk_2} (-1)^{P+P'} \la \bq_2| 
S_{\bk_2} \ra \la \bq'_{N-2}{|\bk'_{N-2}}\ra, \nn \\
{\rm with}~ \la \bq_{2}| S_{\bk_2}\ra &=& \frac{\phi_{\bq_2}^*(\b0)
\phi_{\bk_2} (\b0)}{(1/U - K_{E_{\bk_2}} ({\b0})) (E_{\bk_2}-E_{\bq_2}+i \e)}.
\label{npart} \eea
Here $\bq_2$ ($\bk_2$) 
denotes a pair of momenta chosen from the set $\bqn$ ($\bkn$), and $\bq'$ 
($\bk'$) denotes the remaining $N-2$ momenta. $P$ ($P'$) are the appropriate 
number of permutations. Using Eq. (\ref{npart}), we can calculate the current 
expectation value for the state $|\psi_{\bkn} \ra$ to order $U^2$.
(At order $U^2$, there are also contributions to the current from three- 
and four-particle scattering, but we will ignore those here). 
The current in the incident state $|\phi_{\bkn}\ra$ is given by $\la \phi | 
j_x | \phi \ra = 2 [\sum_{j=1}^N \sin (k_j)] {\cal N}^{N-1}$. The correct
normalization is obtained by dividing by a factor ${\cal{N}}^N$
which then gives in the continuum limit: 
$j_{inc} = [\int_0^{k_L} dk ~2 \sin (k) -\int_0^{k_R} dk~ 2 \sin (k)]/(2\pi)
= (\mu_L-\mu_R)/(2 \pi)$, 
where $k_{L,R}=\cos^{-1} (- \mu_{L,R}/2)$, and we have used $dk=dE/|dE/dk|=
dE/|2 \sin (k)|$. Inserting factors of $\hbar$ and $e$, this 
gives the expected Landauer current $I = (e/h) (\mu_L - \mu_R)$ and Landauer 
conductance $G = e^2 /h$. The change in the Landauer current due to 
two-particle scattering 
is given by a sum of two-particle currents from all possible momentum 
pairs: $\delta j_N= (1/2) \sum_{r,s} \delta j( k_r,k_s) ~{\cal{N}}^{N-2}$ 
which, with the same normalization as used earlier, gives
\bea \delta j_N ~=~ \f{1}{2 (2 \pi)^2} \int \int d k_1 d k_2 ~\delta 
j(k_1, k_2), \label{deljn} \eea
where the integrations are over the full range of allowed momenta $[-\pi ,
\pi]$, and $\delta j(k_1,k_2)$ is given by Eq. (\ref{c2}) [expanded to order 
$U^2$]. Using the fact that $\delta j(k_1,k_2)$ vanishes whenever $k_1,k_2$ 
have opposite signs and converting Eq. (\ref{deljn}) to energy integrals, we 
find the following correction to the Landauer current,
\bea \delta j_N &=& [\int_{-2}^{\mu_R} dE_{k_1} \int_{\mu_R}^{\mu_L} 
dE_{k_2} + \frac{1}{2} \int_{\mu_R}^{\mu_L} dE_{k_1} \int_{\mu_R}^{\mu_L} 
dE_{k_2}] \nn \\
& \times & \rho (E_{k_1}) \rho (E_{k_2}) ~U^2 4 |\phi_{k_1,k_2} (\b0)|^2 ~
Im [K_{E_{k_1,k_2}}(\b0)], \nn \\
& & \label{curr} \eea
where $\rho(E) = 1/(2\pi \sqrt{4-E^2})$ is the density of states. The quantity
in Eq. (\ref{curr}) is negative because $Im [K_{E_{\bk}}(\b0)] < 0$ for all 
values of $\bk$. In the zero bias limit $\mu_L \to \mu_R$, Eq. (\ref{curr}) 
vanishes as $U^2 (\mu_L - \mu_R)$ due to the contribution coming from the first
set of integrals; thus $G$ is less than $e^2 /h$ by a term of order $U^2$.
 

Finally, let us briefly discuss the case of spin-1/2 electrons. We consider 
the Hamiltonian 
$H=-\sum_{x=-\infty}^{\infty} \sum_{\s=\ua ,\da}~ (c_{x,\s}^\dg 
c_{x+1,\s} + h.c.) + U n_{0\ua} n_{0\da}$. 
The interaction at the site 0 can cause scattering between two electrons
in the singlet channel but not in the triplet channel. The scattering
of two electrons in the singlet channel can be studied exactly using the 
Lippman-Schwinger formalism just as in Eqs. (\ref{lipp}-\ref{scatt}), except
that the wave function for the state $|\phi_{\bk} \ra \equiv |k_1, \ua; k_2,
\da \ra = - |k_2, \da; k_1, \ua \ra$ is now given by $\phi_{\bk} (\bl) = 
e^{i(k_1 x_1 + k_2 x_2)}$, and the Green's function is given by
$K_{E_\bk}(\bl) = [1/(2 \pi)^2] \int_{-\pi}^\pi \int_{-\pi}^\pi
{dq_1 dq_2} {e^{i {\bf q} \cdot \bl }}/ (~E_\bk - E_{\bf q} +i \e)$.
Finally, we can argue as in the spinless case, that in the presence of a Fermi
sea, the scattering reduces the Landauer conductance by a term of order $U^2$.

{\bf Discussion:} We have shown how the Lippman-Schwinger formalism can be 
used to obtain exact results for two particles scattering from an interacting 
region. This method can be applied to other cases, such as the two-wire system
studied in Refs. \cite{goorden07}, the case of spin-$1/2$ electrons as 
mentioned above and the case with interactions on more than one bond. We 
have demonstrated how scattering 
theory can be used to understand numerical results for a two-particle wave 
packet moving through the interacting region. Finally, we have considered the 
problem of many-particle transport across the interacting region; we find that
two-particle scattering reduces the zero-temperature Landauer conductance by 
a term of order $U^2$. This calculation is nontrivial since it considers 
many-particle states and is a fully nonequilibrium treatment. We expect the 
two-particle scattering approximation to be valid at low densities 
$k_{L,R}/\pi << 1$ \cite{rech} since the $3$-particle correction given by 
$\int\int\int dk_1dk_2dk_3 \delta j(k_1,k_2,k_3)$ would be smaller by a
factor of order $k_{L,R}/\pi$. In this paper we have considered the
simplest case with interactions on a single bond and no impurities. For 
interactions on more than one bond, the form of the two-particle $S$-matrix 
would change but the qualitative conclusions remain the same \cite{sen}. In 
the presence of impurities however, a term of $O(U)$ appears in the 
correction to $G$ and this could lead to an enhancement of $G$, depending on 
the sign of $U$. More generally, interactions can lead to dephasing and 
suppression of weak localization thereby increasing $G$ \cite{whitney}.


We thank Natan Andrei, Markus B\"uttiker, Leonid Levitov, and Sumathi Rao for
stimulating discussions. 

\end{document}